\documentclass[aps,prd,preprint,superscriptaddress,showpacs,nofootinbib]{revtex4}
\usepackage{graphicx}

\newcommand{\bmat}{\left(\begin{array}}
\newcommand{\emat}{\end{array}\right)}
\newcommand{\beq}{\begin{equation}}
\newcommand{\eeq}{\end{equation}}

\def\bwt{\begin{widetext}}
\def\ewt{\end{widetext}}
\def\be{\begin{equation}}
\def\ee{\end{equation}}
\def\bea{\begin{eqnarray}}
\def\eea{\end{eqnarray}}
\def\bean{\begin{eqnarray*}}
\def\eean{\end{eqnarray*}}
\def\bary{\begin{array}}
\def\eary{\end{array}}
\def\bit{\begin{itemize}}
\def\eit{\end{itemize}}

\def\su5u1{SU(5) \times U(1)}
\def\fsu5u1{SU(5) \times U(1)'}
\def\so10{SO(10)}
\def\sq20{SO(10) \times SO(10)}

\usepackage[centertags]{amsmath}
\usepackage{amssymb}
\newcommand{\Z}{{\mathbb Z}}

\begin{document}

\title{Variations of the Hidden Sector in a Realistic \\Intersecting
Brane Model}

\author{Ching-Ming Chen}

\affiliation{George P. and Cynthia W. Mitchell Institute for
Fundamental Physics,
 Texas A$\&$M University, College Station, TX 77843, USA }

\author{Tianjun Li}

\affiliation{George P. and Cynthia W. Mitchell Institute for
Fundamental Physics, Texas A$\&$M University, College Station, TX
77843, USA }

\affiliation{ Institute of Theoretical Physics, Chinese Academy of
Sciences, Beijing 100080, China}

\author{V. E. Mayes}

\affiliation{George P. and Cynthia W. Mitchell Institute for
Fundamental Physics, Texas A$\&$M University, College Station, TX
77843, USA }

\author{Dimitri V. Nanopoulos}

\affiliation{George P. and Cynthia W. Mitchell Institute for
Fundamental Physics,
 Texas A$\&$M University, College Station, TX 77843, USA }

\affiliation{Astroparticle Physics Group,
Houston Advanced Research Center (HARC),
Mitchell Campus, Woodlands, TX 77381, USA}

\affiliation{Academy of Athens, Division of Natural Sciences,
 28 Panepistimiou Avenue, Athens 10679, Greece }

\date{\today}

\begin{abstract}

Recently, we discussed the first example of a phenomenologically
realistic intersecting D6-brane model.  In this model, the gauge symmetry in the hidden sector
is $USp(2)_1 \times USp(2)_2 \times USp(2)_3 \times USp(2)_4$.
However, we find that the $USp(2)_1\times USp(2)_2$
gauge symmetry can be replaced by an $U(2)_{12}$ gauge symmetry,
and/or the $USp(2)_3\times USp(2)_4$ gauge symmetry
can be replaced by an $U(2)_{34}$ gauge symmetry since
the $USp(2)^2$ stacks of D6-branes contribute to the same
Ramond-Ramond tadpoles as those of the $U(2)$ stacks.
Thus, there are three non-equivalent variations of
the hidden sector, and the corresponding gauge symmetries
are $U(2)_{12} \times USp(2)_3 \times USp(2)_4$,
$U(2)_{34} \times USp(2)_1 \times USp(2)_2$,
and $U(2)_{12} \times U(2)_{34}$, respectively.
Moreover, we study the hidden sector gauge symmetry
breaking, discuss how to decouple the additional
exotic particles, and briefly comment
on the phenomenological consequences.

\end{abstract}


\preprint{ACT-02-07, MIFP-07-11}

\maketitle

\section{Introduction}

The goal of string phenomenology is to construct realistic
standard-like string models with all moduli stabilized. In the
early days, string model building was mainly concentrated on the
weakly coupled heterotic string theory. After the second string
revolution, consistent four-dimensional chiral models with
non-Abelian gauge symmetry on Type II orientifolds were able to be
constructed due to the advent of  D-branes~\cite{JPEW}. In
particular, Type II orientifolds with intersecting D-branes, where
the chiral fermions arise from the intersections of D-branes in
the internal space~\cite{bdl} with T-dual description in terms of
magnetized D-branes~\cite{bachas}, have played an important role
in string model building during the last few years.

On Type IIA orientifolds with intersecting D6-branes,
many non-supersymmetric three-family
standard-like models and Grand Unified Theories (GUTs) were
constructed~\cite{Blumenhagen:2000wh,Angelantonj:2000hi,Blumenhagen:2005mu}.
Although these models were globally consistent, there generically existed
uncancelled Neveu-Schwarz-Neveu-Schwarz (NSNS) tadpoles
as well as the gauge hierarchy problem. To solve these two problems,
semi-realistic supersymmetric standard-like models,
Pati-Salam models, $SU(5)$ models as well as
 flipped $SU(5)$ models have been constructed in
Type IIA theory on $\mathbf{T^6/(\Z_2\times
\Z_2)}$~\cite{CSU1,CSU2,Cvetic:2002pj,CP,CLL,
Cvetic:2004nk,Chen:2005ab,Chen:2005mj} and
$\mathbf{T^6/(\Z_2\times
\Z_2')}$~\cite{Blumenhagen:2005tn,Chen:2006sd} orientifolds with
intersecting D6-branes, and some of their phenomenological
consequences have been studied~\cite{CLS1,CLW}. Moreover, the
supersymmetric constructions in Type IIA theory on other
orientifolds were also discussed~\cite{ListSUSYOthers}. There are
two main constraints on supersymmetric D6-brane model building: RR
tadpole cancellation conditions  and four-dimensional $N=1$
supersymmetric D6-brane configurations. Also, K-theory conditions
provide minor constraints. In addition, to stabilize the
closed-string moduli via supergravity fluxes, the flux models on
Type II orientifolds have also been
constructed~\cite{Blumenhagen:2003vr,Cascales:2003zp,MS,CL,Cvetic:2005bn,
Kumar:2005hf,Chen:2005cf,Villadoro:2005cu,Camara:2005dc,Chen:2006gd,Chen:2006ip}.

It is well known that there are two serious problems in almost all
the supersymmetric D-brane models: no gauge coupling unification
at the string scale, and the rank one problem in the Standard
Model (SM) fermion Yukawa matrices. Although these problems can be
solved in the flux models of Ref.~\cite{Chen:2006gd} where the RR
tadpole cancellation conditions are relaxed, these models are in
the AdS vacua and the question of how to lift these AdS vacua to
the Minkowski vacua or dS vacua correctly is still a big
challenge. Recently, we found that there is  one and only one
intersecting D6-brane model on Type IIA $\mathbf{T^6/(\Z_2\times
\Z_2)}$ orientifold where the above problems can be
solved~\cite{CLL,Chen:2006gd}. Moreover, this model may has a
realistic low energy phenomenology~\cite{Chen:2007px}. Although
its observable sector has unique phenomological properties, it is
possible to have different stacks of the D6-branes in the hidden
sector.

In this paper, we discuss three non-equivalent variations of the
hidden sector where the RR tadpoles are cancelled, the
four-dimensional $N=1$ supersymmetry is perserved, and the
K-theory conditions are satisfied. These three variations seem to
be the only possibilities. In the original
model~\cite{CLL,Chen:2006gd}, the gauge symmetry in the hidden
sector is $USp(2)_1\times USp(2)_2\times USp(2)_3 \times
USp(2)_4$. Interestingly, we can replace the $USp(2)_1\times
USp(2)_2$ gauge symmetry by an $U(2)_{12}$ gauge symmetry, and/or
the $USp(2)_3\times USp(2)_4$ gauge symmetry by an $U(2)_{34}$
gauge symmetry since the contributions to the RR tadpoles from the
$USp(2)^2$ stacks of D6-branes are the same as those of the $U(2)$
stacks. Thus, there are three non-equivalent variations, and the
corresponding gauge symmetries in the hidden sector are $U(2)_{12}
\times USp(2)_3 \times USp(2)_4$, $U(2)_{34} \times USp(2)_1
\times USp(2)_2$, and $U(2)_{12} \times U(2)_{34}$, respectively.
Moreover, we discuss the hidden sector gauge symmetry breaking,
and consider how to decouple the additional exotic particles.
Because the observable sector is the same, the discussions on
phenomenological consequences, for example, the gauge coupling
unification, supersymmetry breaking soft terms, low energy
supersymmetric particle spectrum, dark matter density, and the SM
fermion masses and mixings, are the same as those in
Ref.~\cite{Chen:2007px,CLMN-L}.

This paper is organized as follows. We  briefly review  the
intersecting D6-brane model building on Type IIA
$\mathbf{T^6/(\Z_2\times \Z_2)}$ orientifold in Section II and
the realistic intersecting D6-brane model in Section III. We study
the three variations of the hidden sector in Section IV.
Discussion and conclusions are given in Section V.

\section{Intersecting D6-Brane Model Building in
Type IIA Theory on $\mathbf{T^6/(\Z_2\times \Z_2)}$
Orientifold}

We briefly review the intersecting D6-brane model
building in Type IIA theory on
$\mathbf{T^6/(\Z_2\times \Z_2)}$  orientifold~\cite{CSU1,CSU2}.
 We consider $\mathbf{T^{6}}$ to be a
six torus factorized as
$\mathbf{T^{6}} = {\bf T}^{2} \times {\bf T}^{2} \times {\bf T}^{2}$
whose complex coordinates are $z_i$, $i=1,\; 2,\; 3$ for the
$i$-th two torus, respectively. The $\theta$ and $\omega$
generators for the orbifold group $\Z_{2} \times \Z_{2}$
 act on the complex coordinates as following
\begin{eqnarray}
& \theta: & (z_1,z_2,z_3) \to (-z_1,-z_2,z_3)~,~ \nonumber \\
& \omega: & (z_1,z_2,z_3) \to (z_1,-z_2,-z_3)~.~\,
\label{Z2Z2}
\end{eqnarray}
We implement an orientifold projection $\Omega R$, where $\Omega$
is the world-sheet parity, and $R$ acts on the complex coordinates as
\begin{equation}
R:(z_1,z_2,z_3)\rightarrow(\overline{z}_1,\overline{z}_2,\overline{z}_3)~.~\,
\end{equation}
So, there are four kinds of
orientifold 6-planes (O6-planes) for the actions of $\Omega R$,
$\Omega R\theta$, $\Omega R \omega$, and $\Omega R\theta\omega$,
respectively. Also, we have two kinds of complex structures
consistent with orientifold projection for a two torus --
rectangular and tilted~\cite{LUII}. If we denote the
homology classes of the three cycles wrapped by the D6-brane
stacks as $n_a^i[a_i]+m_a^i[b_i]$ and $n_a^i[a'_i]+m_a^i[b_i]$
with $[a_i']=[a_i]+\frac{1}{2}[b_i]$ for the rectangular and
tilted tori respectively, we can label a generic one cycle by
$(n_a^i,l_a^i)$ in either case, where in terms of the wrapping
numbers $l_{a}^{i}\equiv m_{a}^{i}$ for a rectangular two torus
and $l_{a}^{i}\equiv 2\tilde{m}_{a}^{i}=2m_{a}^{i}+n_{a}^{i}$ for
a tilted two torus.   So, the homology three-cycles for stack $a$
of $N_a$ D6-branes and its orientifold image $a'$ take the form
\beq
[\Pi_a]=\prod_{i=1}^{3}\left(n_{a}^{i}[a_i]+2^{-\beta_i}l_{a}^{i}[b_i]\right),\;\;\;
\left[\Pi_{a'}\right]=\prod_{i=1}^{3}
\left(n_{a}^{i}[a_i]-2^{-\beta_i}l_{a}^{i}[b_i]\right)~,~\, \eeq
where $\beta_i=0$ if the $i$-th two torus is rectangular and
$\beta_i=1$ if it is tilted. Also, we define
$k\equiv\beta_1+\beta_2+\beta_3$.

\begin{table}[t]
\caption{General spectrum for intersecting D6-branes at generic
angles, where $I_{aa'}=-2^{3-k}\prod_{i=1}^3(n_a^il_a^i)$,
and $I_{aO6}=2^{3-k}(-l_a^1l_a^2l_a^3
+l_a^1n_a^2n_a^3+n_a^1l_a^2n_a^3+n_a^1n_a^2l_a^3)$.
Moreover,
${\cal M}$ is the multiplicity, and $a_S$ and $a_A$ denote
 the symmetric and anti-symmetric representations of
$U(N_a/2)$, respectively.}
\renewcommand{\arraystretch}{1.25}
\begin{center}
\begin{tabular}{|c|c|}
\hline {\bf Sector} & \phantom{more space inside this box}{\bf
Representation}
\phantom{more space inside this box} \\
\hline\hline
$aa$   & $U(N_a/2)$ vector multiplet  and 3 adjoint chiral multiplets  \\
\hline
$ab+ba$   & $ {\cal M}(\frac{N_a}{2},
\frac{\overline{N_b}}{2})=
I_{ab}=2^{-k}\prod_{i=1}^3(n_a^il_b^i-n_b^il_a^i)$ \\
\hline
$ab'+b'a$ & $ {\cal M}(\frac{N_a}{2},
\frac{N_b}{2})=I_{ab'}=-2^{-k}\prod_{i=1}^3(n_{a}^il_b^i+n_b^il_a^i)$ \\
\hline $aa'+a'a$ &  ${\cal M} (a_S)=
\frac 12 (I_{aa'} - \frac 12 I_{aO6})$~;~~ ${\cal M} (a_A)=
\frac 12 (I_{aa'} + \frac 12 I_{aO6}) $ \\
\hline
\end{tabular}
\end{center}
\label{spectrum}
\end{table}

For a stack of $N$ D6-branes that do not lie on the top of any
O6-plane, we obtain the $U(N/2)$ gauge symmetry with three adjoint
chiral superfields due to the orbifold projections. While for a
stack of $N$ D6-branes on the top of an O6-plane, we obtain the
$USp(N)$ gauge symmetry with three anti-symmetric chiral
superfields. The bifundamental chiral superfields arise from the
intersections of two different stacks of D6-branes or
 one stack of D6-branes and its $\Omega R$ image~\cite{CSU1,CSU2}.
In short, the general spectrum for intersecting D6-branes at
generic angles, which is valid for both rectangular and tilted two
tori, is given in Table \ref{spectrum}. Moreover, a model may
contain additional non-chiral (vector-like) multiplet pairs from
$ab+ba$, $ab'+b'a$, and $aa'+a'a$ sectors if two stacks of the
corresponding D-branes are parallel and on the top of each other
on one two torus. The multiplicity of the non-chiral multiplet
pairs is given by the product of the intersection numbers on the
other two two-tori.

Before further discussions, let us define the products of wrapping
numbers
\beq
\begin{array}{rrrr}
A_a \equiv -n_a^1n_a^2n_a^3, & B_a \equiv n_a^1l_a^2l_a^3,
& C_a \equiv l_a^1n_a^2l_a^3, & D_a \equiv l_a^1l_a^2n_a^3, \\
\tilde{A}_a \equiv -l_a^1l_a^2l_a^3, & \tilde{B}_a \equiv
l_a^1n_a^2n_a^3, & \tilde{C}_a \equiv n_a^1l_a^2n_a^3, &
\tilde{D}_a \equiv n_a^1n_a^2l_a^3.\,
\end{array}
\label{variables}
\eeq

The four-dimensional $N=1$ supersymmetric models from Type IIA
orientifolds with intersecting D6-branes are mainly constrained by
the RR tadpole cancellation conditions and the four-dimensional
$N=1$ supersymmetric D6-brane configurations, and also
constrained by the K-theory conditions: \\

(1) RR Tadpole Cancellation Conditions \\

The total RR charges of D6-branes and O6-planes must vanish since
the RR field flux lines are conserved. And then we obtain
 the RR tadpole cancellation conditions  as follows
\begin{eqnarray}
 -2^k N^{(1)}+\sum_a N_a A_a=-2^k N^{(2)}+\sum_a N_a
B_a= \nonumber\\ -2^k N^{(3)}+\sum_a N_a C_a=-2^k N^{(4)}+\sum_a
N_a D_a=-16,\,
\end{eqnarray}
where $2 N^{(i)}$ are the number of D6-branes wrapping along
the $i$-th O6-plane which is defined in Table \ref{orientifold}. \\

\renewcommand{\arraystretch}{1.4}
\begin{table}[t]
\caption{Wrapping numbers of the four O6-planes.} \vspace{0.4cm}
\begin{center}
\begin{tabular}{|c|c|c|}
\hline
  Orientifold Action & O6-Plane & $(n^1,l^1)\times (n^2,l^2)\times
(n^3,l^3)$\\
\hline
    $\Omega R$& 1 & $(2^{\beta_1},0)\times (2^{\beta_2},0)\times
(2^{\beta_3},0)$ \\
\hline
    $\Omega R\omega$& 2& $(2^{\beta_1},0)\times (0,-2^{\beta_2})\times
(0,2^{\beta_3})$ \\
\hline
    $\Omega R\theta\omega$& 3 & $(0,-2^{\beta_1})\times
(2^{\beta_2},0)\times
(0,2^{\beta_3})$ \\
\hline
    $\Omega R\theta$& 4 & $(0,-2^{\beta_1})\times (0,2^{\beta_2})\times
    (2^{\beta_3},0)$ \\
\hline
\end{tabular}
\end{center}
\label{orientifold}
\end{table}

(2) Four-Dimensional $N = 1$ Supersymmetric D6-Brane Configurations \\

 The four-dimensional $N=1$ supersymmetry can be preserved by the
orientation projection if and only if the rotation angle of any
D6-brane with respect to the O6-plane is an element of
$SU(3)$~\cite{bdl}, or in other words,
$\theta_1+\theta_2+\theta_3=0$ mod $2\pi$, where $\theta_i$ is the
angle between the D6-brane and the O6-plane in the $i$-th two
torus. This supersymmetry  conditions can be rewritten
as~\cite{Cvetic:2002pj}
\begin{eqnarray}
x_A\tilde{A}_a+x_B\tilde{B}_a+x_C\tilde{C}_a+x_D\tilde{D}_a=0,
\nonumber\\\nonumber \\ A_a/x_A+B_a/x_B+C_a/x_C+D_a/x_D<0,
\label{susyconditions}
\end{eqnarray} where $x_A=\lambda,\;
x_B=\lambda 2^{\beta_2+\beta3}/\chi_2\chi_3,\; x_C=\lambda
2^{\beta_1+\beta3}/\chi_1\chi_3,\; x_D=\lambda
2^{\beta_1+\beta2}/\chi_1\chi_2$, and $\chi_i=R^2_i/R^1_i$ are the
complex structure parameters. The positive parameter $\lambda$ has
been introduced to put all the variables $A,\,B,\,C,~{\rm and}~D$ on an
equal footing.  \\

(3) K-theory Conditions \\

The discrete D-brane RR charges classified by
the $\mathbf{\Z_2}$ K-theory groups in the
presence of orientifolds, which are subtle and invisible by the
ordinary homology~\cite{MS,Witten9810188},
should also be taken into account~\cite{Cascales:2003zp}.
The K-theory conditions are
\begin{eqnarray}
\sum_a 2^{-k}\tilde{A}_a  = \sum_a 2^{-\beta_1} N_a  \tilde{B}_a
= \sum_a 2^{-\beta_2} N_a  \tilde{C}_a =
\sum_a 2^{-\beta_3} N_a  \tilde{D}_a
 = 0 \textrm{ mod }4 \label{K-charges}~.~\,
\end{eqnarray}

\section{The Realistic Intersecting D6-Brane Model}

There may be one and only one intersecting D6-brane model in
Type IIA theory on $\mathbf{T^6/(\Z_2\times \Z_2)}$ orientifold
with a realistic phenomenology~\cite{CLL,Chen:2006gd,Chen:2007px}.
Let us briefly review it. We present the D6-brane
configurations and intersection numbers in
Table~\ref{MI-Numbers}, and its spectrum in Table~\ref{Spectrum}.
We put the $a'$, $b$, and $c$ stacks of D6-branes on the top of
each other on the third two torus, and then we have the additional
vector-like particles from $N=2$ subsectors.

\begin{table}[h]

\begin{center}

\footnotesize

\begin{tabular}{|@{}c@{}|c||@{}c@{}c@{}c@{}||c|c||c|@{}c@{}|@{}c@{}|
@{}c@{}||@{}c@{}|@{}c@{}|@{}c@{}|@{}c@{}|} \hline

stack & $N$ & ($n_1$,$l_1$) & ($n_2$,$l_2$) & ($n_3$,$l_3$) & A &
S & $b$ & $b'$ & $c$ & $c'$ & $O6^{1}$ & $O6^{2}$ & $O6^{3}$ &
$O6^{4}$ \\ \hline \hline

$a$ & 8 & ( 0,-1) & ( 1, 1) & ( 1, 1) & 0 & 0 & 3 & 0(3) & -3 &
0(3) & 1 & -1 & 0 & 0 \\ \hline

$b$ & 4 & ( 3, 1) & ( 1, 0) & ( 1,-1) & -2 & 2 & - & - & 0(6) &
0(1) & 0 & 1 & 0 & -3  \\  \hline

$c$ & 4 & ( 3,-1) & ( 0, 1) & ( 1,-1) & 2 & -2 & - & - & - & - &
-1 & 0 & 3 & 0 \\ \hline \hline

$O6^{1}$ & 2 & ( 1, 0) & ( 1, 0) & ( 2, 0) & - & - & - & - & - & -
& - & - & - & -  \\ \hline

$O6^{2}$ & 2 & ( 1, 0) & ( 0,-1) & ( 0, 2) & - & - & - & - & - & -
& - & - & - & -  \\ \hline

$O6^{3}$ & 2 & ( 0, -1) & ( 1, 0) & ( 0, 2) & - & - & - & - & - &
- & - & - & - & -  \\ \hline

$O6^{4}$ & 2 & ( 0, -1) & ( 0, 1) & ( 2, 0) & - & - & - & - & - &
- & - & - & - & -  \\ \hline
\end{tabular}
\caption{The D6-brane configurations and intersection numbers on Type
IIA $\mathbf{T}^6 / \Z_2 \times \Z_2$ orientifold. The gauge
symmetry is $[U(4)_C \times U(2)_L \times
U(2)_R]_{\rm Observable}\times [USp(2)_1 \times USp(2)_2
\times USp(2)_3 \times USp(2)_4]_{\rm Hidden}$, the SM fermions
and Higgs fields arise from the intersections on the first
two torus, and the complex structure parameters are
$2\chi_1=6\chi_2=3\chi_3 =6$. Also, the beta functions for all
$USp(2)_i$ gauge symmetries are $-3$.}
\label{MI-Numbers}
\end{center}
\end{table}

\begin{table}
[htb] \footnotesize
\renewcommand{\arraystretch}{1.0}
\caption{The chiral and vector-like superfields,
 and their quantum numbers
under the gauge symmetry $SU(4)_C\times SU(2)_L\times SU(2)_R
\times USp(2)_1 \times USp(2)_2 \times USp(2)_3 \times USp(2)_4$.}
\label{Spectrum}
\begin{center}
\begin{tabular}{|c||c||c|c|c||c|c|c|}\hline
 & Quantum Number
& $Q_4$ & $Q_{2L}$ & $Q_{2R}$  & Field \\
\hline\hline
$ab$ & $3 \times (4,\overline{2},1,1,1,1,1)$ & 1 & -1 & 0  & $F_L(Q_L, L_L)$\\
$ac$ & $3\times (\overline{4},1,2,1,1,1,1)$ & -1 & 0 & $1$   & $F_R(Q_R, L_R)$\\
$a1$ & $1\times (4,1,1,2,1,1,1)$ & $1$ & 0 & 0  & $X_{a1}$ \\
$a2$ & $1\times (\overline{4},1,1,1,2,1,1)$ & -1 & 0 & 0   & $X_{a2}$ \\
$b2$ & $1\times(1,2,1,1,2,1,1)$ & 0 & 1 & 0    & $X_{b2}$ \\
$b4$ & $3\times(1,\overline{2},1,1,1,1,2)$ & 0 & -1 & 0    & $X_{b4}^i$ \\
$c1$ & $1\times(1,1,\overline{2},2,1,1,1)$ & 0 & 0 & -1    & $X_{c1}$ \\
$c3$ & $3\times(1,1,2,1,1,2,1)$ & 0 & 0 & 1   &  $X_{c3}^i$ \\
$b_{S}$ & $2\times(1,3,1,1,1,1,1)$ & 0 & 2 & 0   &  $T_L^i$ \\
$b_{A}$ & $2\times(1,\overline{1},1,1,1,1,1)$ & 0 & -2 & 0   & $S_L^i$ \\
$c_{S}$ & $2\times(1,1,\overline{3},1,1,1,1)$ & 0 & 0 & -2   & $T_R^i$  \\
$c_{A}$ & $2\times(1,1,1,1,1,1,1)$ & 0 & 0 & 2   & $S_R^i$ \\
\hline\hline
$ab'$ & $3 \times (4,2,1,1,1,1,1)$ & 1 & 1 & 0  & \\
& $3 \times (\overline{4},\overline{2},1,1,1,1,1)$ & -1 & -1 & 0  & \\
\hline
$ac'$ & $3 \times (4,1,2,1,1,1,1)$ & 1 &  & 1  & $\Phi_i$ \\
& $3 \times (\overline{4}, 1, \overline{2},1,1,1,1)$ & -1 & 0 & -1  &
$\overline{\Phi}_i$\\
\hline
$bc$ & $6 \times (1,2,\overline{2},1,1,1,1)$ & 0 & 1 & -1   & $H_u^i$, $H_d^i$\\
& $6 \times (1,\overline{2},2,1,1,1,1)$ & 0 & -1 & 1   & \\
\hline
\end{tabular}
\end{center}
\end{table}

We have shown that the gauge symmetry in the observable sector can
be broken down to the SM gauge symmetry via the Green-Schwarz mechanism,
 D6-brane splittings and supersymmtry preserving Higgs mechanism.
The gauge couplings for $SU(4)_C$, $SU(2)_L$ and $SU(2)_R$ are
unified at the string scale, and the additional exotic particles
may be decoupled around the string scale. Also, we calculated the
supersymmetry breaking soft terms, and the corresponding low
energy supersymmetric particle spectrum that can be tested at the
Large Hadron Collider (LHC). The observed dark matter density can
also be generated. In addition, we can explain the SM quark masses
and mixings, and the tau lepton mass. The neutrino masses and
mixings may be generated via seesaw mechanism as well. Similar to
the GUTs~\cite{Nanopoulos:1982zm}, we have roughly the wrong
fermion mass relation $m_e/m_{\mu} \simeq m_{d}/m_s$, and the
correct electron and muon masses can be generated via
high-dimensional operators~\cite{CLMN-L}. Furthermore, all the
$USp(2)_i$ gauge symmetries will become strong around the string
scale~\cite{CLMN-L}.

\section{Three variations of the Hidden Sector}

In the realistic intersecting D6-brane model~\cite{CLL,Chen:2006gd},
the observable sector is unique.
Interestingly, we find three non-equivalent variations of
the hidden sector where we can cancel the RR tadpoles,
 preserve the  four-dimensional $N=1$ supersymmetry,
and satisfy the K-theory conditions. And it seems to us that there
is no other variation.  In the original
model~\cite{CLL,Chen:2006gd}, the gauge symmetry in the hidden
sector is $USp(2)_1\times USp(2)_2\times USp(2)_3 \times
USp(2)_4$. We notice that the $USp(2)_1\times USp(2)_2$ gauge
symmetry can be replaced by an $U(2)_{12}$ gauge symmetry, and/or
the $USp(2)_3\times USp(2)_4$ gauge symmetry by an $U(2)_{34}$
gauge symmetry because the contributions to the RR tadpoles from
the $USp(2)^2$ stacks of D6-branes are the same as those of the
$U(2)$ stacks. Thus, there are three non-equivalent variations,
and the corresponding gauge symmetries in the hidden sector are
$U(2)_{12} \times USp(2)_3 \times USp(2)_4$, $U(2)_{34} \times
USp(2)_1 \times USp(2)_2$, and $U(2)_{12} \times U(2)_{34}$,
respectively. Let us present them one by one in the following
subsections.

\subsection{$U(2)_{12}\times USp(2)_3 \times USp(2)_4$ Hidden Sector}

\begin{table}[h]
\begin{center}
\footnotesize
\begin{tabular}{|@{}c@{}|c||@{}c@{}c@{}c@{}||c|c||c|@{}c@{}|@{}c@{}|
@{}c@{}||@{}c@{}|@{}c@{}|@{}c@{}|@{}c@{}|} \hline

stack & $N$ & ($n_1$,$l_1$) & ($n_2$,$l_2$) & ($n_3$,$l_3$) & A &
S & $b$ & $b'$ & $c$ & $c'$ & $d$ & $d'$ & $O6^{3}$ & $O6^{4}$
\\ \hline \hline

$a$ & 8 & ( 0,-1) & ( 1, 1) & ( 1, 1) & 0 & 0 & 3 & 0(3) & -3 &
0(3) & 0(2) & 0(1) & 0 & 0 \\ \hline

$b$ & 4 & ( 3, 1) & ( 1, 0) & ( 1,-1) & -2 & 2 & - & - & 0(6) &
0(1) & 1 & 0(1) & 0 & -3  \\  \hline

$c$ & 4 & ( 3,-1) & ( 0, 1) & ( 1,-1) & 2 & -2 & - & - & - & - &
-1 & 0(1) & 3 & 0 \\ \hline \hline

$d$ & 4 & ( 1, 0) & ( 1,-1) & ( 1, 1) & 0 & 0 & - & - & - & - & -
& - & -1 & 1  \\ \hline

$O6^{3}$ & 2 & ( 0, -1) & ( 1, 0) & ( 0, 2) & - & - & - & - & - &
- & - & - & - & -  \\ \hline

$O6^{4}$ & 2 & ( 0, -1) & ( 0, 1) & ( 2, 0) & - & - & - & - & - &
- & - & - & - & -  \\ \hline
\end{tabular}
\caption{The D6-brane configurations and intersection numbers on Type
IIA $\mathbf{T}^6 / \Z_2 \times \Z_2$ orientifold. The complete
gauge symmetry is $[U(4)_C \times U(2)_L \times
U(2)_R]_{\rm Observable}\times [U(2)_{12} \times
USp(2)_3\times USp(2)_4]_{\rm Hidden}$.}
\label{HA-MI-Numbers}
\end{center}
\end{table}

\begin{table}[htb]
\footnotesize
\renewcommand{\arraystretch}{1.0}
\caption{The new chiral superfields and their quantum numbers
under the gauge symmetry $SU(4)_C\times SU(2)_L\times
SU(2)_R \times U(2)_{12} \times USp(2)_3 \times USp(2)_4 $.}

\label{HA-Spectrum}
\begin{center}
\begin{tabular}{|c||c||c|c|c|c||c|c|c|}\hline

& Representation

& $Q_4$ & $Q_{2L}$ & $Q_{2R}$ & $Q_{12} $ & Field \\

\hline\hline

$bd$ & $1\times(1,2,1,\overline{2},1,1)$ & 0 & 1 & 0 & -1   & $X_{bd}$ \\

$cd$ & $1\times(1,1,\overline{2},2,1,1)$ & 0 & 0 & -1 & 1  &  $X_{cd}$ \\

$d3$ & $1\times (1,1,1,\overline{2},2,1)$ & 0 & 0 & 0  & -1 & $X_{d3}$ \\

$d4$ & $1\times (1,1,1,2,1,2)$ & 0 & 0 & 0   & 1 & $X_{d4}$ \\
\hline

\end{tabular}
\end{center}
\end{table}

In the first variation of the hidden sector, we replace
the $USp(2)_1\times USp(2)_2$ gauge symmetry by an
$U(2)_{12}$ gauge symmetry.  We present the D6-brane
configurations and intersection numbers in
Table~\ref{HA-MI-Numbers}. Moreover, the particle spectrum has two parts:
(1) the spectrum for old particles is given
in Table~\ref{Spectrum} by removing all the particles that
are charged under $USp(2)_1\times USp(2)_2$; (2) the spectrum
for the  new particles is given in Table~\ref{HA-Spectrum}.

The anomalies from the global $U(1)$ of $U(2)_{12}$ are cancelled
by the Green-Schwarz mechanism, and its gauge field obtains mass
via the linear $B\wedge F$ couplings. Then, the effective gauge
symmetry is $SU(2)_{12}$.  The $SU(2)_{12}$ gauge symmetry can be
broken down to $U(1)_{12}$ via D6-brane splitting. Interestingly,
we do not have any additional chiral exotic particles that are
charged under $SU(4)_C$. The simple way to give masses to the
extra exotic particles $X_{bd}$ and $X_{cd}$ is instanton
effects~\cite{Blumenhagen:2006xt,Ibanez:2006da,Cvetic:2007ku,Ibanez:2007rs}.
However, we do not have the suitable three-cycles wrapped by E2
instantons~\footnote{Note that the E2 branes must also wrap rigid
cycles.}, and thus the instanton effects are not available.
Similar results hold for the next two subsections.
 In addition, the $USp(2)_3$ and $USp(2)_4$ will become strong
at about the string scale~\cite{CLMN-L}, and then we will have
some composite particles in the $U(2)_{12}$ anti-symmetric and
symmetric  representations, $\overline{S}'_d$ and
$\overline{T}'_d$ from $X_{d3}$, and $S'_d$ and $T'_d$ from
$X_{d4}$, respectively.  So we can break the $U(1)_{12}$ by giving
suitable string-scale vacuum expectation values (VEVs) to
$\overline{T}'_d$ and $T'_d$, and we can give the string-scale
VEVs to  $\overline{S}'_d$ and $S'_d$.  Note that we give the
TeV-scale VEVs to $S_L^i$ and the string-scale VEVs to
$S_R^i$~\cite{Chen:2007px}, we can give the GUT-scale masses to
$X^i_{c3}$ and $X_{cd}$ and the TeV-scale masses to the $X^i_{b4}$
and $X_{bd}$ via the high-dimensional operators~\cite{CLMN-L}.
Furthermore,  if we could give the string-scale masses to the
three $U(2)_{12}$ adjoint chiral superfields and we do not break
the $SU(2)_{12}$ via D6-brane splitting, the $SU(2)_{12}$ gauge
symmetry will become strong around the string scale. Then we can
have the singlet composite field $S'_L$ in the $U(2)_L$
anti-symmetric representation with charge $+2$ under $U(1)_L$ from
$X_{bd}$. And we can give the string-scale VEVs to $S_L^i$ and
$S'_L$ while keeping the D-flatness of $U(1)_L$. Therefore, we may
also give the GUT-scale masses to the $X^i_{b4}$ and $X_{bd}$ via
the high-dimensional operators~\cite{CLMN-L}.

\subsection{$U(2)_{34}\times USp(2)_1 \times USp(2)_2$ Hidden Sector}

\begin{table}[h]

\begin{center}

\footnotesize

\begin{tabular}{|@{}c@{}|c||@{}c@{}c@{}c@{}||c|c||c|@{}c@{}|@{}c@{}|
@{}c@{}||@{}c@{}|@{}c@{}|@{}c@{}|@{}c@{}|} \hline

stack & $N$ & ($n_1$,$l_1$) & ($n_2$,$l_2$) & ($n_3$,$l_3$) & A &
S & $b$ & $b'$ & $c$ & $c'$ & $e$ & $e'$ & $O6^{1}$ & $O6^{2}$
\\ \hline \hline

$a$ & 8 & ( 0,-1) & ( 1, 1) & ( 1, 1) & 0 & 0 & 3 & 0(3) & -3 &
0(3) & 0(2) & 0(0) & 1 & -1 \\ \hline

$b$ & 4 & ( 3, 1) & ( 1, 0) & ( 1,-1) & -2 & 2 & - & - & 0(6) &
0(1) & 0(3) & -3 & 0 & 1  \\  \hline

$c$ & 4 & ( 3,-1) & ( 0, 1) & ( 1,-1) & 2 & -2 & - & - & - & - &
0(3) & 3 & -1 & 0 \\ \hline \hline

$e$ & 4 & ( 0, 1) & (-1, 1) & (-1, 1) & 0 & 0 & - & - & - & - & -
& - & -1 & 1 \\ \hline

$O6^{1}$ & 2 & ( 1, 0) & ( 1, 0) & ( 2, 0) & - & - & - & - & - & -
& - & - & - & -  \\ \hline

$O6^{2}$ & 2 & ( 1, 0) & ( 0,-1) & ( 0, 2) & - & - & - & - & - & -
& - & - & - & -  \\ \hline

\end{tabular}
\caption{The D6-brane configurations and intersection numbers on Type
IIA $\mathbf{T}^6 / \Z_2 \times \Z_2$ orientifold.
The gauge symmetry is $[U(4)_C \times
U(2)_L \times U(2)_R]_{\rm Observable}\times [U(2)_{34}\times
USp(2)_1\times USp(2)_2]_{\rm Hidden}$. }
\label{HB-MI-Numbers}
\end{center}
\end{table}

\begin{table}[htb]
\footnotesize
\renewcommand{\arraystretch}{1.0}
\caption{The new chiral superfields and their quantum numbers
under the gauge symmetry $SU(4)_C\times SU(2)_L\times
SU(2)_R \times U(2)_{34} \times USp(2)_1 \times USp(2)_2 $.}

\label{HB-Spectrum}
\begin{center}
\begin{tabular}{|c||c||c|c|c|c||c|c|c|}\hline

& Representation

& $Q_4$ & $Q_{2L}$ & $Q_{2R}$ & $Q_{34}$ & Field \\

\hline\hline

$be'$ & $3\times(1,\overline{2},1,\overline{2},1,1)$ & 0 & -1 & 0  & -1  & $X^i_{be'}$ \\

$ce'$ & $3\times(1,1,2,2,1,1)$ & 0 & 0 & 1 & 1  & $X^i_{ce'}$ \\

$e1$ & $1\times (1,1,1,\overline{2},2,1)$ & 0 & 0 & 0  & -1 & $X_{e1}$ \\

$e2$ & $1\times (1,1,1,2,1,2)$ & 0 & 0 & 0   & 1 & $X_{e2}$ \\

\hline

\end{tabular}
\end{center}
\end{table}

In the second variation of the hidden sector, we replace the
$USp(2)_3\times USp(2)_4$ gauge symmetry by an $U(2)_{34}$ gauge
symmetry.  We present the D6-brane configurations and intersection
numbers in Table~\ref{HB-MI-Numbers}. The particle spectrum also
has two parts: (1) the spectrum for old particles is given in
Table~\ref{Spectrum} by removing all the particles that are
charged under $USp(2)_3\times USp(2)_4$; (2) the spectrum for the
new particles is given in Table~\ref{HB-Spectrum}.

Note that the wrapping numbers for the $d$ stack of D6-branes are
equivalent to those of the $a$ stack by T duality and orientifold
action, we can think that we have an $U(6)$ gauge symmetry in the
begining. Only the global $U(1)$ of $U(6)$ is anomalous $U(1)$ symmetry,
and its gauge field obtains mass via the linear $B\wedge F$ couplings.
After we put four D6-branes on the place with equivalent wrapping
numbers (just like the D6-brane splittings), we break the $SU(6)$ down to the
$SU(4)_C \times SU(2)_{34} \times U(1)'$ where the $U(1)'$ generator
in $SU(6)$ is
\begin{eqnarray}
 T_{U(1)'}
\equiv ~ {1\over {2{\sqrt 6}}} \, {\rm diag}\left(1, 1, 1, 1, -2, -2 \right) ~.~\,
\label{SU6-GU1A}
\end{eqnarray}
Thus, the left-handed and right-handed SM fermions have $U(1)'$ charges
$+1/2{\sqrt 6}$ and $-1/2{\sqrt 6}$, respectively.
In order to keep the gauge coupling
unification, we have to break the $U(1)'$ so that it will not become part
of the $U(1)_Y$. In short, we have to break $U(2)_{34}$
completely.

Because the $USp(2)_1$ and $USp(2)_2$ will become strong at about
the string scale~\cite{CLMN-L}, we will have some composite
particles in the $U(2)_{34}$ anti-symmetric and symmetric
representations, $\overline{S}'_e$ and $\overline{T}'_e$ from
$X_{e1}$, and $S'_e$ and $T'_e$ from $X_{e2}$, respectively. So we
can break the $U(2)_{12}$ completely by giving suitable
string-scale VEVs to $\overline{S}'_e$, $\overline{T}'_e$, $S'_e$,
and $T'_e$. Moreover, we can have the singlet composite particle
$S'_L$ in the $U(2)_L $ anti-symmetric representation with charge
$+2$ under $U(1)_L$ from $X_{b2}$. And then we can give the
string-scale VEVs to $S_L^i$ and $S'_L$ while keeping the
D-flatness of $U(1)_L$. Note that $S_R^i$ also have string-scale
VEVs, we may give the GUT-scale masses to $X_{b2}$, $X_{c1}$,
$X^i_{be'}$, and $X^i_{ce'}$ via the high-dimensional
operators~\cite{CLMN-L}. Moreover, $X_{a1}$ and $X_{a2}$ may form
the vector-like particles if we break the $USp(2)_1$ and
$USp(2)_2$ down to the diagonal $USp(2)_{D12}$~\cite{Chen:2007px}.

\subsection{$U(2)_{12}\times U(2)_{34}$ Hidden Sector}

\begin{table}[h]
\begin{center}
\footnotesize
\begin{tabular}{|@{}c@{}|c||@{}c@{}c@{}c@{}||c|c||c|@{}c@{}|@{}c@{}|
@{}c@{}||@{}c@{}|@{}c@{}|@{}c@{}|@{}c@{}|} \hline

stack & $N$ & ($n_1$,$l_1$) & ($n_2$,$l_2$) & ($n_3$,$l_3$) & A &
S & $b$ & $b'$ & $c$ & $c'$ & $d$ & $d'$ & $e$ & $e'$
\\ \hline \hline

$a$ & 8 & ( 0,-1) & ( 1, 1) & ( 1, 1) & 0 & 0 & 3 & 0(3) & -3 &
0(3) & 0(2) & 0(1) & 0(2) & 0(0) \\ \hline

$b$ & 4 & ( 3, 1) & ( 1, 0) & ( 1,-1) & -2 & 2 & - & - & 0(6) &
0(1) & 1 & 0(1) & 0(3) & -3  \\  \hline

$c$ & 4 & ( 3,-1) & ( 0, 1) & ( 1,-1) & 2 & -2 & - & - & - & - &
-1 & 0(1) & 0(3) & 3 \\ \hline \hline

$d$ & 4 & ( 1, 0) & ( 1,-1) & ( 1, 1) & 0 & 0 & - & - & - & - & -
& - & 0(1) & 0(2)  \\ \hline

$e$ & 4 & ( 0, 1) & (-1, 1) & (-1, 1) & 0 & 0 & - & - & - & - & -
& - & - & - \\ \hline

\end{tabular}
\caption{The D6-brane configurations and intersection numbers on
Type IIA $\mathbf{T}^6 / \Z_2 \times \Z_2$ orientifold. The
complete gauge symmetry is $[U(4)_C \times U(2)_L \times
U(2)_R]_{\rm Observable}\times [U(2)_{12}\times U(2)_{34}]_{\rm Hidden}$. }
\label{HC-MI-Numbers}
\end{center}
\end{table}

\begin{table}[htb]
\footnotesize
\renewcommand{\arraystretch}{1.0}
\caption{The chiral and vector-like superfields,
 and their quantum numbers under the gauge symmetry
 $SU(4)_C\times SU(2)_L\times
SU(2)_R \times U(2)_{12}\times U(2)_{34}$.}
\label{HC-Spectrum}
\begin{center}
\begin{tabular}{|c||c||c|c|c|c|c||c|c|c|}\hline

& Representation

& $Q_4$ & $Q_{2L}$ & $Q_{2R}$ & $Q_{12}$ & $Q_{34}$ & Field \\

\hline\hline

$bd$ & $1\times(1,2,1,\overline{2},1)$ & 0 & 1 & 0 & -1 & 0   & $X_{bd}$ \\

$be'$ & $3\times(1,\overline{2},1,1,\overline{2})$ & 0 & -1 & 0 & 0 & -1  & $X_{be'}^i$ \\

$cd$ & $1\times(1,1,\overline{2},2,1)$ & 0 & 0 & -1 & 1 & 0  &  $X_{cd}$ \\

$ce'$ & $3\times(1,1,2,1,2)$ & 0 & 0 & 1 & 0 & 1   & $X_{ce'}^i$ \\

\hline\hline

$de$ & $1 \times (1,1,1,2,\overline{2})$ & 0 & 0 & 0 & 1 & -1 & $X_{de}$ \\
& $1 \times (1,1,1,\overline{2},2)$ & 0 & 0 & 0 & -1 & 1  & $\overline{X}_{de}$ \\

\hline

$de'$ & $2 \times (1,1,1,2,2)$ & 0 & 0 & 0 & 1 & 1  &  $X_{de'}^i$ \\
& $2 \times (1,1,1,\overline{2},\overline{2})$ & 0 & 0 & 0 & -1 & -1  &
$\overline{X}_{de'}^i$ \\

\hline

\hline
\end{tabular}
\end{center}
\end{table}

In the third variation of the hidden sector, we replace the
$USp(2)_1\times USp(2)_2$ gauge symmetry by $U(2)_{12}$, and
replace the $USp(2)_3\times USp(2)_4$ gauge symmetry by
$U(2)_{34}$.  We present the D6-brane configurations and
intersection numbers in Table~\ref{HC-MI-Numbers}. The particle
spectrum also has two parts: (1) the spectrum for old particles is
given in Table~\ref{Spectrum} by removing all the particles that
are charged under $USp(2)_1\times USp(2)_2\times USp(2)_3\times
USp(2)_4$; (2) the spectrum for the  new particles is given in
Table~\ref{HC-Spectrum}.

As discussed in above subsections, we can break the $U(2)_{12}$
down to the $U(1)_{12}$ gauge symmetry via Green-Schwarz mechanism
and D6-brane splitting, and we have to break the $U(2)_{34}$
gauge symmetry completely. In order to break the $U(1)_{12}$ and
$U(2)_{34}$ gauge symmetries, we put the $d$ and $e$ stacks of
D6-branes on the top of each other on the second two torus, and
put the $d$ and $e'$ stacks on the top of each other on the third
two torus. Then, we have additional vector-like particles
$X_{de}$, $\overline{X}_{de}$, $X^i_{de'}$, and
$\overline{X}^i_{de'}$, as given in Table~\ref{HC-Spectrum}. And
there exist the following Yukawa couplings
\begin{eqnarray}
W \supset && y^A_{ij} X_{bd} X_{be'}^i X_{de'}^j +
y^B_{ij} X_{cd} X_{ce'}^i \overline{X}_{de'}^j~,~\,
\end{eqnarray}
where $y^A_{ij}$ and $y^B_{ij}$ are Yukawa couplings. If we give
the diagonal string-scale VEVs to $X_{de'}^j$ and
$\overline{X}_{de'}^j$, we break the $U(2)_{12}\times U(2)_{34}$
down to the diagonal $U(2)_D$. Moreover, the $X_{bd}$ and one
linear combination of $X_{be'}^i$, and the $X_{cd}$ and one linear
combination of $X_{ce'}^i$ can have vector-like masses close to
the string scale. Note that we can give the TeV-scale VEVs to
$S_L^i$ and the string-scale VEVs to $S_R^i$~\cite{Chen:2007px},
we can give the GUT-scale masses to $X_{cd}$ and the other two
linear combinaions of $X_{ce'}^i$, and the TeV-scale masses to $X_{bd}$
and the other two linear combinations of $X_{be'}^i$ via the
high-dimensional operators~\cite{CLMN-L}. Similar to the
discussions in the above subsection A,  if we can give the
string-scale masses to the three $U(2)_{12}$ adjoint chiral
superfields and do not break the $SU(2)_{12}$ gauge symmetry via
D6-brane splitting, the $SU(2)_{12}$ gauge symmetry will become
strong around the string scale. Then we can have the singlet
composite field $S'_L$ in the $U(2)_L$ anti-symmetric
representation with charge $+2$ under $U(1)_L$ from $X_{bd}$, and
we can give the string-scale VEVs to $S_L^i$ and $S'_L$ while
keeping the D-flatness of $U(1)_L$. Therefore, we may also give
the GUT-scale masses to $X_{bd}$ and the other two linear combinations of
$X_{be'}^i$ via the high-dimensional operators~\cite{CLMN-L}.

\section{Discussion and Conclusions}

At present, there is only one known example of an intersecting
D6-brane model with a realistic observable sector.  Interestingly,
there are three non-equivalent variations of the hidden sector in
which the theoretical constraints on model building can be
satisfied. There does not seem to be any other possible variation
in the original model~\cite{CLL,Chen:2006gd}, and the gauge symmetry
in the hidden sector is $USp(2)_1\times USp(2)_2\times USp(2)_3
\times USp(2)_4$. We noticed that the $USp(2)_1\times USp(2)_2$
gauge symmetry can be replaced by an $U(2)_{12}$ gauge symmetry,
and/or the $USp(2)_3\times USp(2)_4$ gauge symmetry can be
replaced by an $U(2)_{34}$ gauge symmetry because the $USp(2)^2$
stacks of D6-branes contribute to the same RR tadpoles as those of
the $U(2)$ stacks. Thus, we obtained three non-equivalent
variations, and the corresponding gauge symmetries in the hidden
sector are $U(2)_{12} \times USp(2)_3 \times USp(2)_4$, $U(2)_{34}
\times USp(2)_1 \times USp(2)_2$, and $U(2)_{12} \times
U(2)_{34}$, respectively. In addition, we studied the hidden
sector gauge symmetry breaking, and discussed how to decouple the
additional exotic particles. Because the observable sector is the
same, the phenomenological discussions in the observable sector
are the same as those in Ref.~\cite{Chen:2007px,CLMN-L}.

\section*{Acknowledgments}
This research was supported in part
by the Mitchell-Heep Chair in High Energy Physics (CMC),
by the Cambridge-Mitchell Collaboration in Theoretical Cosmology (TL),
and by the DOE grant DE-FG03-95-Er-40917 (DVN).


\begin{thebibliography}{99}
\itemsep 0.5mm


\bibitem{JPEW}
J.~Polchinski and E.~Witten, Nucl.\ Phys.\ B {\bf 460}, 525
(1996).

\bibitem{bdl}
M.~Berkooz, M.~R.~Douglas and R.~G.~Leigh, Nucl. Phys. B {\bf 480}
(1996) 265.

\bibitem{bachas}
C.~Bachas, arXiv:hep-th/9503030.


\bibitem{Blumenhagen:2000wh}
  R.~Blumenhagen, L.~Goerlich, B.~Kors and D.~Lust,
  JHEP {\bf 0010}, 006 (2000).


\bibitem{Angelantonj:2000hi}
  C.~Angelantonj, I.~Antoniadis, E.~Dudas and A.~Sagnotti,
  Phys.\ Lett.\ B {\bf 489}, 223 (2000).



\bibitem{Blumenhagen:2005mu}
  R.~Blumenhagen, M.~Cvetic, P.~Langacker and G.~Shiu,
  Ann.\ Rev.\ Nucl.\ Part.\ Sci.\  {\bf 55}, 71 (2005), and the references therein.




\bibitem{CSU1}
M.~Cveti\v c, G.~Shiu and A.~M.~Uranga, Phys.\ Rev.\ Lett.\  {\bf
87}, 201801 (2001).

\bibitem{CSU2}
M.~Cveti\v c, G.~Shiu and A.~M.~Uranga, Nucl.\ Phys.\ B {\bf 615},
3 (2001).


\bibitem{Cvetic:2002pj}
  M.~Cveti\v c, I.~Papadimitriou and G.~Shiu,
  Nucl.\ Phys.\ B {\bf 659}, 193 (2003)
  [Erratum-ibid.\ B {\bf 696}, 298 (2004)].


\bibitem{CP} M. Cveti\v c and I. Papadimitriou,
Phys.\ Rev.\ D {\bf 67}, 126006 (2003).


\bibitem{CLL}
M.~Cveti\v c, T.~Li and T.~Liu,
Nucl.\ Phys.\ B {\bf 698}, 163 (2004).

\bibitem{Cvetic:2004nk}
  M.~Cveti\v c, P.~Langacker, T.~Li and T.~Liu,
  Nucl.\ Phys.\ B {\bf 709}, 241 (2005).


\bibitem{Chen:2005ab}
  C.-M.~Chen, G.~V.~Kraniotis, V.~E.~Mayes, D.~V.~Nanopoulos and J.~W.~Walker,
  Phys.\ Lett.\ B {\bf 611}, 156 (2005);
Phys.\ Lett.\  B {\bf 625}, 96 (2005).


\bibitem{Chen:2005mj}
  C.-M.~Chen, T.~Li and D.~V.~Nanopoulos,
  Nucl.\ Phys.\ B {\bf 732}, 224 (2006).

\bibitem{Blumenhagen:2005tn}
  R.~Blumenhagen, M.~Cvetic, F.~Marchesano and G.~Shiu,
  JHEP {\bf 0503}, 050 (2005).

\bibitem{Chen:2006sd}
  C.-M.~Chen, V.~E.~Mayes and D.~V.~Nanopoulos,
  arXiv:hep-th/0612087.


\bibitem{CLS1}
 M.~Cveti\v c, P.~Langacker and G.~Shiu,
Phys.\ Rev.\ D {\bf 66}, 066004 (2002);
 Nucl.\ Phys.\ B {\bf 642}, 139 (2002).


\bibitem{CLW}
M.~Cveti\v c, P.~Langacker and J.~Wang,
Phys.\ Rev.\ D {\bf 68}, 046002 (2003).


\bibitem{ListSUSYOthers}
R.~Blumenhagen, L.~G\"orlich and T.~Ott, JHEP {\bf 0301}, 021 (2003);
G.~Honecker, Nucl.\ Phys.\  {\bf B666}, 175 (2003);
G.~Honecker and T.~Ott,
Phys.\ Rev.\ D {\bf 70}, 126010 (2004)
  [Erratum-ibid.\ D {\bf 71}, 069902 (2005)].


\bibitem{Blumenhagen:2003vr}
  R.~Blumenhagen, D.~Lust and T.~R.~Taylor,
  Nucl.\ Phys.\  B {\bf 663}, 319 (2003).


\bibitem{Cascales:2003zp}
  J.~F.~G.~Cascales and A.~M.~Uranga,
  JHEP {\bf 0305}, 011 (2003).


\bibitem{MS}
F.~Marchesano and G.~Shiu,
Phys.\ Rev.\ D {\bf 71}, 011701 (2005);
JHEP {\bf 0411}, 041 (2004).

\bibitem{CL}
M.~Cveti\v c and T.~Liu, Phys.\ Lett.\ B {\bf 610}, 122 (2005).


\bibitem{Cvetic:2005bn}
  M.~Cveti\v c, T.~Li and T.~Liu,
  Phys.\ Rev.\ D {\bf 71}, 106008 (2005).


\bibitem{Kumar:2005hf}
  J.~Kumar and J.~D.~Wells,
  JHEP {\bf 0509}, 067 (2005).


\bibitem{Chen:2005cf}
  C.-M.~Chen, V.~E.~Mayes and D.~V.~Nanopoulos,
  Phys.\ Lett.\  B {\bf 633}, 618 (2006).


\bibitem{Villadoro:2005cu}
  G.~Villadoro and F.~Zwirner,
  JHEP {\bf 0506}, 047 (2005).


\bibitem{Camara:2005dc}
  P.~G.~Camara, A.~Font and L.~E.~Ibanez,
  JHEP {\bf 0509}, 013 (2005).


\bibitem{Chen:2006gd}
  C.-M.~Chen, T.~Li and D.~V.~Nanopoulos,
  Nucl.\ Phys.\  B {\bf 740}, 79 (2006).


\bibitem{Chen:2006ip}
  C.-M.~Chen, T.~Li and D.~V.~Nanopoulos,
  Nucl.\ Phys.\  B {\bf 751}, 260 (2006).

\bibitem{Chen:2007px}
  C.-M.~Chen, T.~Li, V.~E.~Mayes and D.~V.~Nanopoulos,
  arXiv:hep-th/0703280.


\bibitem{LUII}
R.~Blumenhagen, B.~K\"ors and D.~L\"ust, JHEP {\bf 0102} (2001) 030.


\bibitem{Witten9810188}
  E.~Witten,
  JHEP {\bf 9812}, 019 (1998).




\bibitem{Nanopoulos:1982zm}
J.\,R.~Ellis and M.\,K.~Gaillard,
  Phys.\ Lett.\  B {\bf 88}, 315 (1979);
  D.\,V.~Nanopoulos and M.~Srednicki,
  Phys.\ Lett.\  B {\bf 124}, 37 (1983).


\bibitem{CLMN-L}
C.-M.~Chen, T.~Li, V.\,E.~Mayes and D.\,V.~Nanopoulos, in
preparation.


\bibitem{Blumenhagen:2006xt}
  R.~Blumenhagen, M.~Cvetic and T.~Weigand,
  arXiv:hep-th/0609191.

\bibitem{Ibanez:2006da}
  L.~E.~Ibanez and A.~M.~Uranga,
  JHEP {\bf 0703}, 052 (2007).


\bibitem{Cvetic:2007ku}
  M.~Cvetic, R.~Richter and T.~Weigand,
  arXiv:hep-th/0703028.

\bibitem{Ibanez:2007rs}
  L.~E.~Ibanez, A.~N.~Schellekens and A.~M.~Uranga,
  arXiv:0704.1079 [hep-th].


\end{thebibliography}
\end{document}